\documentclass[lettersize,journal]{IEEEtran}
\usepackage{amsmath,amsfonts}
\usepackage{algorithmic}
\usepackage{array}
\usepackage{textcomp}
\usepackage{stfloats}
\usepackage{url}
\usepackage{verbatim}
\usepackage{graphicx}
\usepackage{cite}
\usepackage{subfigure}
\usepackage{booktabs} %三线包
\usepackage{multirow} %多栏包
\usepackage{caption}
\usepackage[inkscapelatex=false]{svg} %SVG包
\def\BibTeX{{\rm B\kern-.05em{\sc i\kern-.025em b}\kern-.08em
    T\kern-.1667em\lower.7ex\hbox{E}\kern-.125emX}}
\usepackage{balance}

\begin{document}
\title{Harnessing Web3 on Carbon Offset Market for Sustainability: Framework and A Case Study}
\author{Chenyu Zhou, Hongzhou Chen, Shiman Wang, Xinyao Sun, Abdulmotaleb El Saddik, Wei Cai

\thanks{This work was supported by the Shenzhen Science and Technology Program (Grant No. JCYJ20210324124205016), in part by the Shenzhen Key Lab of Crowd Intelligence Empowered Low-Carbon Energy Network (Grant No. ZDSYS20220606100601002), in part by the CUHK(SZ)-White Matrix Joint Metaverse Laboratory, in part by the SeeDAO,  and in part by the Shenzhen Institute of Artificial Intelligence and Robotics for Society. \textit{(Corresponding author: Wei Cai.)}}

\thanks{Chenyu Zhou is with the School of Science and Engineering, The Chinese University of Hong Kong, Shenzhen, China. E-mail: chenyuzhou1@link.cuhk.edu.cn.}

\thanks{Hongzhou Chen is with the School of Science and Engineering, The Chinese University of Hong Kong, Shenzhen, China, Mohamed bin Zayed University of Artificial Intelligence, Abu Dhabi, United Arab Emirates, and the Insight and Research Guild, SeeDAO. E-mail: hongzhouchen1@link.cuhk.edu.cn.}% <-this % stops a space

\thanks{Shiman Wang is with the Insight and Research Guild, SeeDAO. E-mail: shiman272@outlook.com.}

\thanks{Xinyao Sun is with Matrix Labs Inc., Vancouver, BC, Canada. E-mail: asun@matrixlabs.org.}

\thanks{Abdulmotaleb El Saddik is with Mohamed bin Zayed University of Artificial Intelligence, Abu Dhabi, United Arab Emirates, and the Multimedia Communication Research Laboratory (MCRLab), University of Ottawa, Ottawa, Canada. E-mail: elsaddik@uottawa.ca.}
\thanks{Wei Cai is with the School of Science and Engineering, The Chinese University of Hong Kong, Shenzhen, China, and Shenzhen Institute of Artificial Intelligence and Robotics for Society, Shenzhen, China. E-mail: caiwei@cuhk.edu.cn.}

}

\markboth{Journal of \LaTeX\ Class Files,~Vol.~18, No.~9, September~2020}%
{How to Use the IEEEtran \LaTeX \ Templates}

\maketitle

\begin{abstract}

Blockchain, pivotal in shaping the metaverse and Web3, often draws criticism for high energy consumption and carbon emission. The rise of sustainability-focused blockchains, especially when intersecting with innovative wireless technologies, revises this predicament. To understand blockchain's role in sustainability, we propose a three-layers structure encapsulating four green utilities: Recording and Tracking, Wide Verification, Value Trading, and Concept Disseminating. Nori, a decentralized voluntary carbon offset project, serves as our case, illuminating these utilities. Our research unveils unique insights into the on-chain carbon market participants, affect factors of the market, value propositions of NFT-based carbon credits, and the role of social media to spread the concept of carbon offset. We argue that blockchain's contribution to sustainability is significant, with carbon offsetting potentially evolving as a new standard within the blockchain sector.

\end{abstract}


\section{Introduction}

\IEEEPARstart{B}{lockchain} technologies underpin user interactions and the ecosystem of the metaverse~\cite{duan2021metaverse}. The application of blockchain technologies, such as smart contract-based tokenomics and non-fungible tokens (NFTs), brings financial incentives and clarifies ownership of digital assets~\cite{wang2021blockchain}, promoting the current Web3 boom. However, blockchain has been criticized for its high energy consumption and carbon emissions. Hence, many blockchain projects like Ethereum upgraded to Proof-of-Stake (PoS) to reduce energy consumption. The PoS upgrade is expected to reduce carbon emissions by more than 99\%~\cite{PKMG2022blockchain}, making blockchain represent the sustainable development of the metaverse.
%~\cite{nair2021evaluation}

\begin{figure*}[]   %To put the Fig.1 into page 2
\centering
\centerline{\includegraphics[width = 18cm]{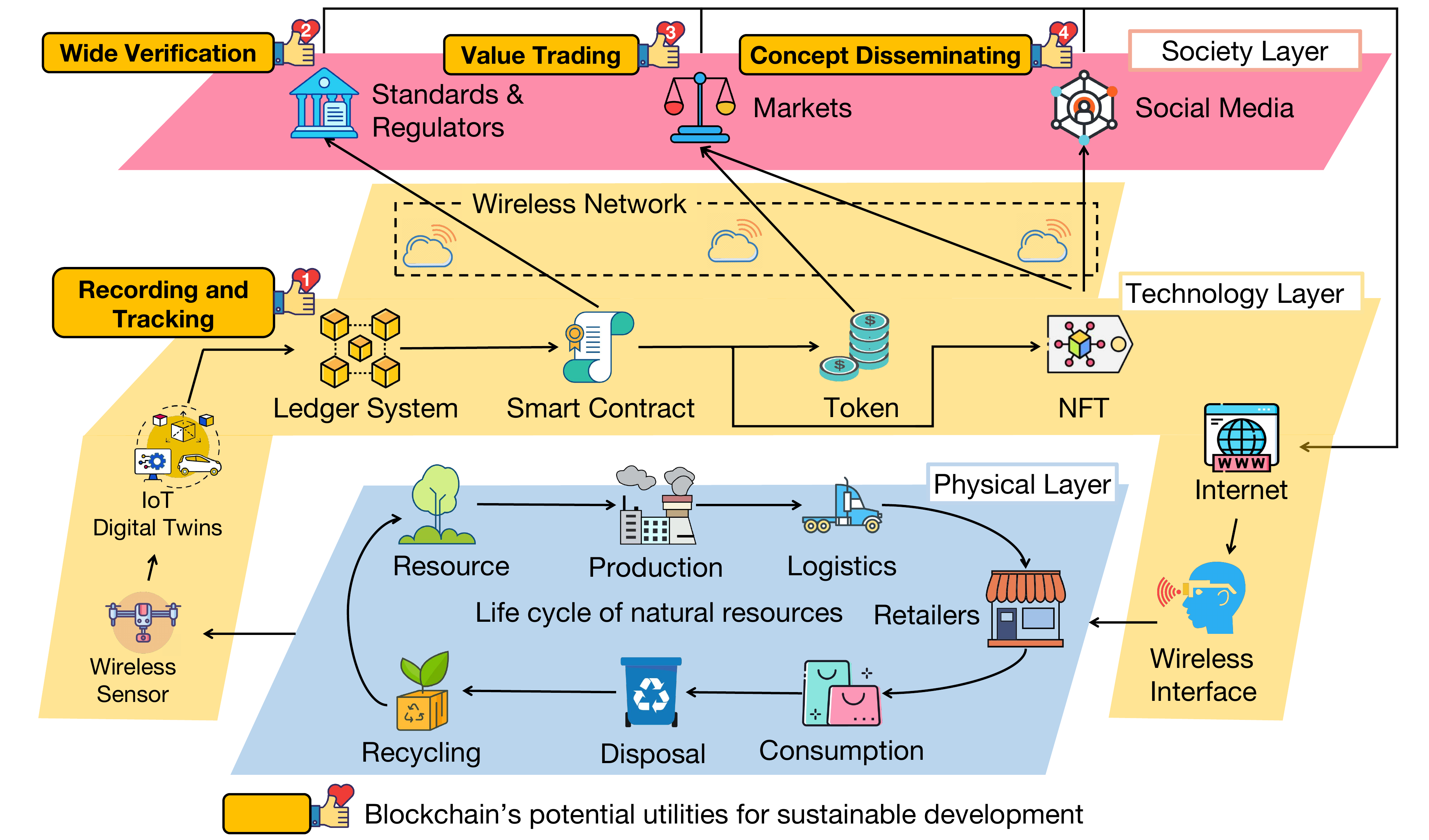}}
%% %^& 这个标题扩充一下
\caption{Three-layers structure of blockchain for sustainability. The physical layer shows the life cycle of natural resources. The second layer shows blockchain and wireless technologies that support sustainable development. The third layer shows relevant societal applications. Specifically, thumbs-up icons reveal the four sustainability utilities.}
\label{fig:structure}
\end{figure*}

% With its transparency and immutability, blockchain can solve dilemmas under a centralized socio-technical architecture. 
Furthermore, intersecting with wireless technologies, blockchain holds further potential for promoting sustainability. In Fig.~\ref{fig:structure}, we propose a three-layer (physical, technological, and societal layer) structure based on the life cycle of natural resources to overview blockchain's four potential utilities for the environment, namely \textbf{Recording and Tracking, Wide Verification, Value Trading, and Concept Disseminating.} At the physical layer, innovative wireless technologies allow for comprehensive and accurate natural resources data collection. At the technological layer, blockchain ledger systems provide a transparent and immutable record of resource transformations, facilitated by innovative networks such as IoT, federated learning, and digital twins~\cite{du2020spacechain, xiaofei2022InFEDge, lv2022building}. At the societal layer, smart contracts can help standards and regulators provide decentralized verification for sustainable entities~\cite{dapp}. Tokens and relevant value trading can incentivize environment-friendly behaviors. And NFTs, recently popular on social media, can promote the dissemination and acceptance of sustainable concepts.

A significant application scenario for Web3's sustainability utilities is the carbon offset market. Carbon offsetting refers to the process of absorbing or reducing carbon emissions through actions such as forest preservation and the use of renewable energy sources, which has two types of market: mandatory and voluntary. Especially, the voluntary market, mainly driven by social responsibility and corporate image~\cite{blaufelder2021blueprint}, reflecting the ongoing global commitment to Environmental, Social, and Governance (ESG) factors. However, the voluntary carbon market struggles with the challenges of centralization, like opaque carbon credit trading and over-reliance on weak regulation. Additionally, high cost, ``double-count'', and high trading thresholds further complicate the market dynamics~\cite{blaufelder2021blueprint}. As a response, Nori\footnote{https://nori.com/}, a Web3 voluntary carbon offset project, employs blockchain-based decentralized ledger technology to record and track carbon credits, issuing tamper-proof carbon offset NFT certificates through smart contracts, and extracting the value of carbon credits, attracting public attention to carbon offset. Through these efforts, Nori has successfully fostered the growth of the voluntary carbon market and exhibited the immense potential of Web3 for sustainability.

Through empirical analysis, we have revealed how Nori uses the socio-technical features of blockchain technology to facilitate the voluntary carbon markets. First, as a Web3 project, Nori attracts a diverse range of carbon credit NFT buyers, including not only crypto-native users like The Sandbox\footnote{https://www.sandbox.game/} and STEPN\footnote{https://stepn.com/} but also traditional companies. This broad alliance amplifies Nori's future market potential by harnessing the strengths of both Web2 and Web3. Second, we have identified four distinct clusters of Nori carbon credit end-users, each characterized by their unique motivations. This insight paints a comprehensive picture of the user profiles active in the voluntary carbon market. Third, our econometric tests suggest that although energy prices do influence Nori's market, it is more responsive to public opinion, indicating a robust and sustained contribution to sustainable development. Finally, our sentiment analysis of Nori-related Twitter discussions, along with eight distinct topic frames, reveal that Nori has gained widespread acceptance and recognition in the public eye.

The contributions of this paper can be concluded as follows,
\begin{itemize}

\item[1)] We proposed a three-layer structure to describe four utilities of Web3 for sustainability.
\item[2)] We have selected Nori to gain insights into the on-chain voluntary carbon market and to delineate the contributions of blockchain.
\item[3)] We analyzed the motivations and behaviors of market participants, market affect factors, and related topic frames to provide clues for future works.  

\end{itemize}

%\section{Blockchain for sustainable development}
\section{Web3 and Wireless for Sustainability}

The intersection of blockchain and wireless technologies holds significant potential. The transparency and immutability inherent in blockchain technology, coupled with the real-time data transfer and broad reach of wireless technology, form a powerful synergy for sustainability. To explore these potentials, we propose a structure that echoes the life cycle of natural resources, embedded with the technology and societal aspects. Since sustainable development engages with natural resources, technology, and social behaviors~\cite{engjellushe2013education}, we divide this structure into three layers: physical, technological, and societal. Fig.~\ref{fig:structure} illustrates this structure, and we highlight four potential utilities that blockchain can contribute to sustainable development, boosted by the inclusion of wireless technology:

\textbf{Recording and Tracking}. Ensuring comprehensive, accurate data records is essential for effectively managing the resource lifecycle, which spans manufacturing, logistics, sales, consumption, disposal, and reclamation. Given the decentralization of these data sources, wireless technologies, such as IoT and sensors, allow for comprehensive and accurate data collection across the lifecycle. Blockchain, serving as a secure and decentralized data storage system, ensures the integrity and transparency of these records, making them accessible to regulators and consumers. By integrating blockchain with wireless technologies~\cite{du2020spacechain, lv2022building}, we can establish a system that promotes responsible resource usage through effective management and traceability.

\textbf{Wide Verification}. Wireless technologies enable the global transfer of real-time data throughout the natural resource lifecycle, and blockchain facilitates the decentralized verification of these data. Traditional approaches, like the Kyoto Protocol's reliance on numerous agencies for verification, pose substantial information synchronization challenges. Comparatively, smart contracts can use data within the decentralized ledger to issue certificates for compliance entities~\cite{hammi2018bctrust}, widely including manufacturers, brands, and even smaller entities, such as vehicles within the logistics chain~\cite{rahman2021blockchain}, significantly reducing the cost of trust and enabling worldwide collaboration.

\textbf{Value Trading}. By incentivizing sustainable practices, blockchain encourages environmental stewardship among individuals and corporations~\cite{dapp}. These contributions to sustainability, such as reducing plastic usage or participating in environmental governance, can be tokenized to produce tradable economic value. Token airdrops can promote initiatives like carbon offsetting, influencing user behavior~\cite{SizhengFMWC2023}. The network effect generated by these tokens increases interest in and commitment to environmental issues, which is based on the deployment of innovative wireless networks.

\textbf{Concept Disseminating}. NFTs guarantee the uniqueness and ownership of digital assets, including authenticating educational degrees~\cite{li2019blockchain}. They can also disseminate information and consensus across industries~\cite{baytacs2022stakeholders}. For instance, NFTs are often associated with artwork, attracting people to share their NFTs on social media and forming relevant virtual communities~\cite{kugler2021non}. Therefore, NFTs can serve as certificates for carbon offsetting, generating more public resonance for sustainability than traditional carbon credit purchases. The widespread accessibility and real-time nature of wireless technologies help for faster dissemination and greater public engagement in sustainable concepts.

\section{Nori, A Web3 Voluntary Carbon Market}

% nori 概述

Nori integrates regenerative agriculture with wireless technologies and blockchain to innovate in the realm of voluntary carbon offsetting. By employing a combination of soil sensors, and utilizing satellites and drones for infrared (IFR) imaging, Nori is able to precisely measure the impact of its own carbon sequestration projects. Additionally, Nori generates the NRTs (Nori Carbon Removal Tonne, a carbon credit NFT deployed on Polygon\footnote{Polygon is a scalable, Ethereum-compatible blockchain platform designed to enable efficient, low-cost transactions for decentralized applications. Its innovative Layer 2 solutions contribute to environmental sustainability by significantly reducing energy consumption and carbon emissions associated with blockchain transactions.}) for every ton of carbon dioxide removed. Buyers can purchase NRTs to offset their carbon emissions. Nori has helped end buyers offset over 125k tons of carbon dioxide and paid over \$2 million to regenerative agriculture developers, while maintaining an average growth rate of 217\% in transaction volume over the past three years. Each NRT is now on sale for an initial price of \$20, and then can be traded at flexible prices in secondary markets.

% 自愿碳市场

\begin{figure}[htbp]
\centering
\centerline{\includegraphics[width = \columnwidth]{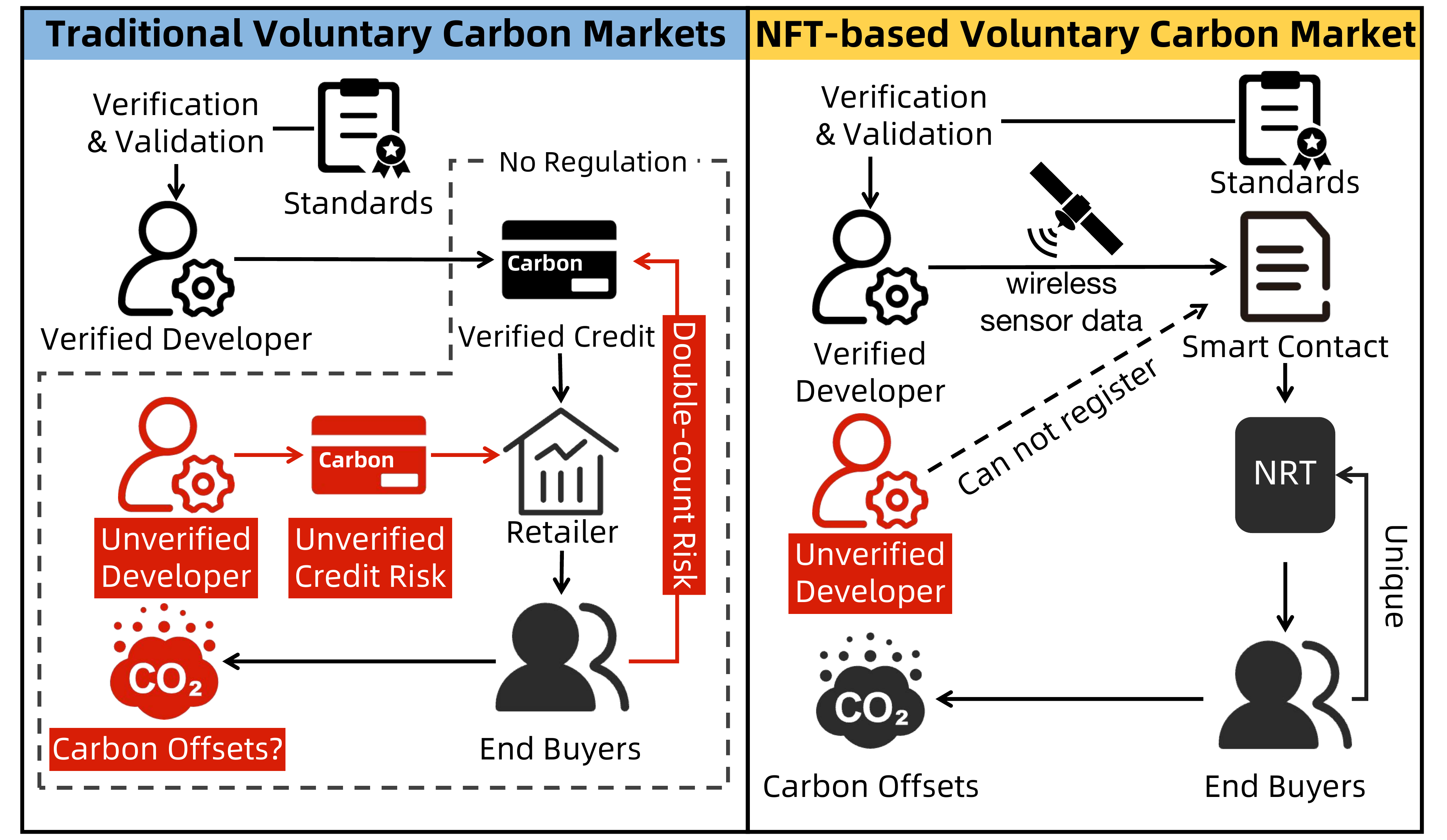}}
\caption{The structures of the traditional voluntary carbon market and Nori NFT-based voluntary carbon market, together with the market's problems and Nori improvements.The red elements are challenges in the traditional market.}
\label{fig:market}
\end{figure}

% \subsection{Voluntary Carbon Offset Market} 
\subsection{Voluntary Carbon Offset Market and Wireless Technology} 

Voluntary carbon offset markets have grown in popularity as a complement to mandatory regulations in line with the Paris Agreement's goal of limiting global temperature increases. Unlike mandatory markets, these emerging markets aim to build a voluntary, wider, and more efficient system. These markets allow end users to purchase carbon credits from developers and retailers to offset their emissions. By facilitating transactions, voluntary carbon markets enable efficient fund transfers and carbon offsetting across companies, industries, and political entities, promoting social good. However, traditional voluntary markets face regulatory challenges, such as investments in non-verified credits or double-counting, which impact carbon offset validity. The structure and issues of the traditional voluntary carbon market are shown in Fig.~\ref{fig:market}.

For carbon sequestration measurement, advanced wireless communication technologies are emerging as indispensable tools. IFR imaging satellites enable real-time monitoring of vegetation and soil, while drone sensor data tracking provides high-resolution insights. Together, these technologies enhance the accuracy and efficiency of measuring carbon capture, contributing to a more equitable and effective operation of the carbon market.

\subsection{Nori and blockchain-based improvements}
Similar to traditional carbon offset markets, Nori registers carbon dioxide absorbed by vegetation photosynthesis as tradable carbon credits. However, by introducing the NRTs representing carbon credits, Nori enhances the efficiency and transparency of the voluntary carbon market. Understanding the function of this blockchain-based mechanism is best achieved by examining the roles of various market participants. Firstly, carbon credit project developers, after being validated by third-party standards, can interact with Nori's smart contract to register the amount of carbon they have legitimately removed. Secondly, end buyers are given the opportunity to directly purchase carbon credits from developers by minting an NRT from Nori's smart contract. Meanwhile, unverified developers, who lack the necessary authorization to register the amount of carbon removed, are prevented from introducing unverified carbon credits into the market. Finally, each carbon credit's uniqueness and the ownership of the end buyer are ensured through NRTs, making double-counting tricks impossible. The structure and improvement of Nori's voluntary carbon market are shown in Fig.~\ref{fig:market}. In terms of data fairness, Nori has been continually iterating its carbon sequestration measurement model through wireless technologies, and by integrating IoT and AI technologies, it strives to achieve carbon sequestration verification with minimal human resource costs.

While Nori holds promise in the carbon market, real-world implementation faces challenges. Regulatory and governance issues may arise due to the fact that many national and regional carbon emission reduction regulations do not recognize carbon offsets purchased from third-party markets. The inherent decentralization of blockchain technology, the foundation for the NRTs, could challenge governments' and regulatory bodies' ability to effectively monitor and control the market. Data privacy and security concerns may emerge due to required disclosures. Traditional carbon market players' resistance may also hinder NRTs adoption, requiring extensive educational initiatives to facilitate the transition.

\begin{figure*}[htbp]
\centering
\centerline{\includegraphics[width = 18cm]{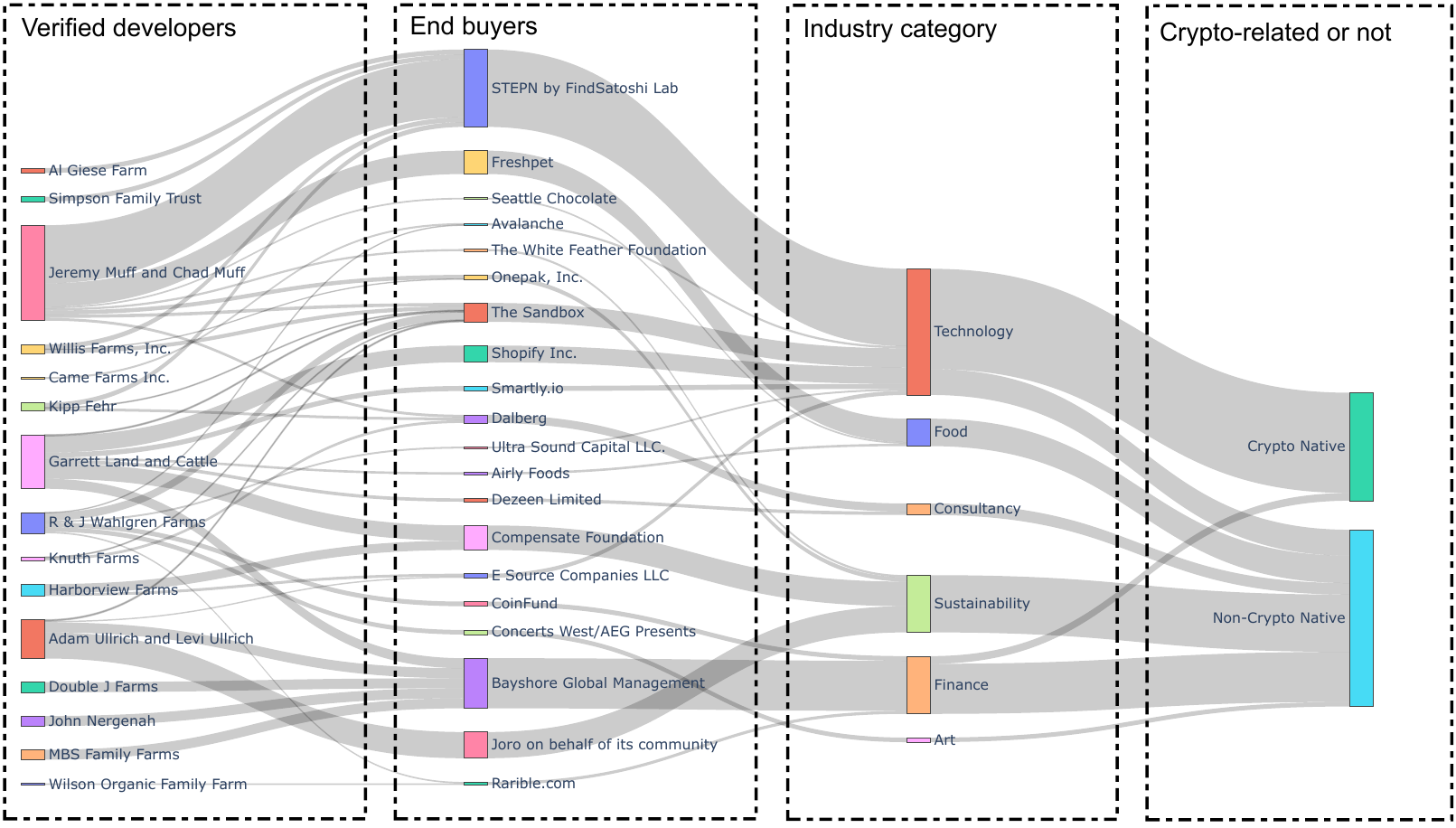}}
% \caption{Transaction flow diagram of end buyers purchasing more than 500 NRTs.}
\caption{Transaction flow diagram of end buyers purchasing more than 500 NRTs. The box on the far left represents certified carbon offset developers, the second box from the left represents buyers, while the right side displays the industry the buyer belongs to, along with the percentage of crypto native companies.}
\label{fig:sankey}
\end{figure*}

\section{Experimental Analysis}

We analyzed the Nori market to understand the utilities of blockchain technology for sustainability. We collected the complete transaction data for Nori NRTs. However, a challenge we face is the ongoing transition of Nori NRTs from the Ethereum (ETH) network to the Polygon network. As a consequence of this transition process, the NRT transaction data may be incomplete. To overcome this issue, we obtained all transaction data from the official NRTs proof website\footnote{https://nori.com/certificate/1} and cross-validated it with the on-chain data. Moreover, since NRTs include the personal/business information of the buyers, we also retrieved this corresponding information from each NFT to construct comprehensive user profiles. To examine the factors influencing the market, we collected data on the public energy and cryptocurrency markets, keyword trend indices from Google Trends, and Twitter text data using web crawling techniques. More detailed introductions to these data will be presented in subsequent subsections.

\subsection{End Buyers source analytics}

After filtering out anonymous transactions, we obtained 2217 records and identified 1516 buyers. Focusing on those purchasing over 500 NRTs, we analyzed major market participants. Based on industry categories and blockchain relevance, we visualized the transaction flow in Fig.~\ref{fig:sankey}. The first column displays verified carbon credit developers who issued NRTs. The second column represents end buyers, and the third column shows their industry categories, revealing where Nori carbon offsets contribute to sustainability. The fourth column classifies buyers as either a crypto-native project or not, indicating whether the project is based on blockchain technology and may have been launched through funding methods such as ICOs (Initial Coin Offering).

The Nori market's main buyers encompass traditional carbon offsets participants, and novel blockchain-related projects, which contribute to carbon offsets from a Web3 perspective. The Top 5 end buyers are: \#1 STEPN, a popular on-chain game focused on outdoor sports and health, aiming for carbon neutrality through Nori. \#2 Bayshore Global Management, an investment firm founded by Google co-founder Sergey Brin and Anne Wojcicki. \#3 Joro, a carbon-neutral startup offering one-stop carbon accounting and offsetting solutions. \#4 Compensate Foundation, a Finnish non-profit addressing various greenhouse events. \#5 Freshpet, a NASDAQ-listed pet food company with annual revenues of \$320 million. The top 5 end buyers data implies that the composition of the primary buyers in the Nori market significantly diverges from that of the main buyers in the mandatory carbon market, which is predominantly made up of energy, steel, chemical, and other heavy industries. Enterprises that are not subject to carbon quota restrictions are actively participating in the Nori market with a voluntary carbon offsetting approach. In addition, the data in the fourth column of the image shows that more than half of the top buyers on Nori are from organizations outside the Web3 area. These features reveal that Nori's market has the potential to overcome the shortages of other decentralized financial projects since it focuses on external physical carbon sink assets and sustainable positive externalities. Moreover, the participation of traditional industries and carbon offset institutions is evidence of the advantages of blockchain-based recording and tracking of carbon credits. 

\subsection{Clustering Analysis of Buyer Behavior}

\begin{figure}[htbp]
\centering
\centerline{\includegraphics[width = \columnwidth]{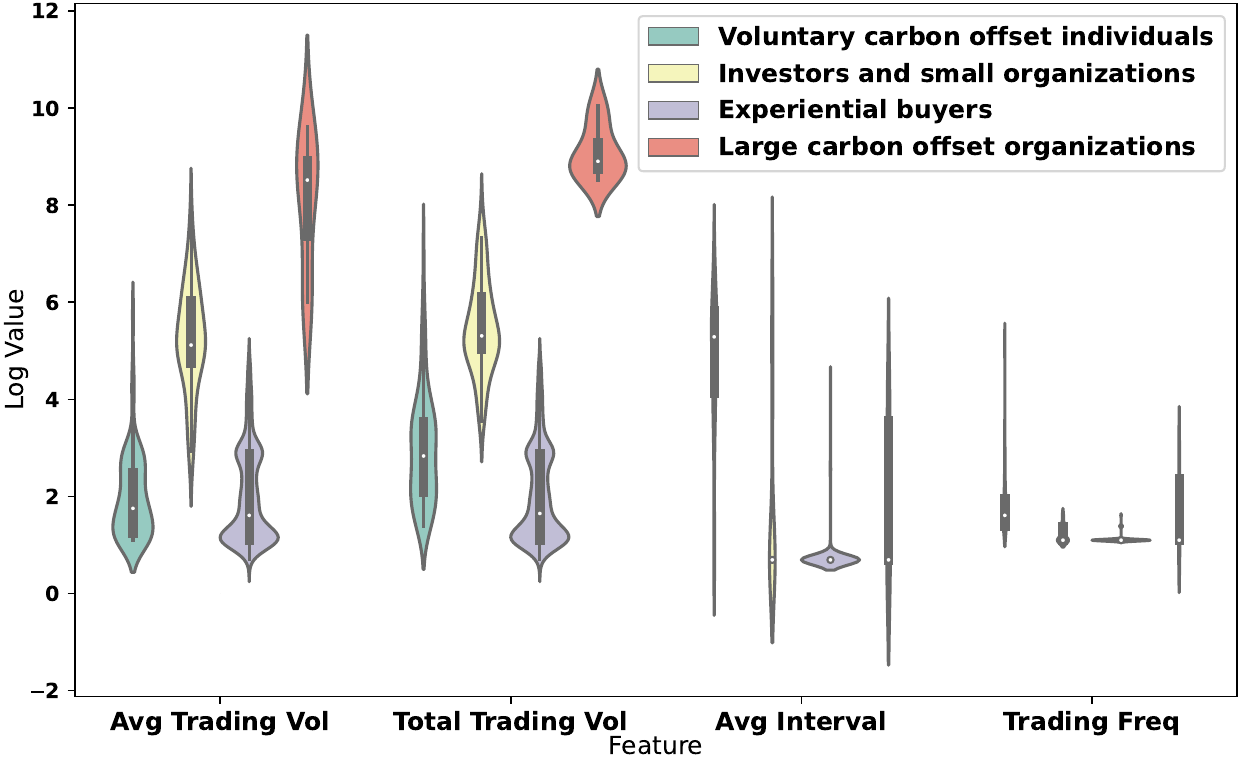}}
\caption{Cluster analysis reveals four distinct buyer categories in the Nori market: Experiential buyers, Voluntary carbon offset individuals, Investors and small organizations, and Large carbon offset organizations.}
\label{fig:cluster_feature}
\end{figure}

We constructed buyer trading behavior features and used an open-source implementation\footnote{https://github.com/JustGlowing/minisom} of Self-Organizing Map (SOM) clustering to examine buyer categories in Nori's market. SOM is an unsupervised learning algorithm that is used for clustering and visualization by organizing data into a grid of nodes, where similar data points are mapped close to each other, and dissimilar data points are mapped further apart. It does this by iteratively adjusting the weights of the nodes to resemble the input data. Buyer features include five dimensions: Total Trading Volume, Average Trading Volume, Average Transaction Interval, Trading Frequency, and Buyer Type (whether the buyer is an individual or a business). After clustering, we identified four categories of end buyers. Fig.~\ref{fig:cluster_feature} displays the statistical analysis of each cluster's characteristics.

\textbf{Cluster 1: Experiential buyers}. Buyers in this cluster are individual buyers who come for project experience or trial investment, accounting for 87.9\% of all buyers. The number and frequency of transactions for this category of buyers are low, mostly one time or one NRT.

\textbf{Cluster 2: Voluntary carbon offset individuals}. This cluster represents low-volume, high-frequency NRT buyers in the market, accounting for 7.6\% of all. They regularly purchase NRTs, with average annual purchases between 4 and 6 tons, aligning with their average annual emissions. This cluster consists of environmentalists practicing personal carbon offsets. For instance, David Dibble recorded six purchases with 14 NRTs over three years, offsetting an average of 4.6 tons of carbon emissions annually.

\textbf{Cluster 3: Investors and Small Organizations}. Representing 4\% of the market, this cluster includes individuals and organizations with analogous trading behavior. Individual buyers mainly consist of investors acquiring substantial quantities of NRTs for potential appreciation or individuals with high carbon liabilities. On the other hand, small and medium-sized enterprises (SMEs) generally make annual purchases of carbon offsets post-emission accounting. For example, Jeff Thiel, an individual buyer, made six purchases totaling 90 tons of carbon offsets over three years, indicating profit-driven investment. A typical SME example is E Source, which bought 1,236 tons of carbon offsets twice at an annual frequency.

\textbf{Cluster 4: Large Carbon Offset Organizations}. This cluster encompasses several large institutions, making up 4\% of the buyers but 58.3\% of market transactions. These entities regularly reconcile their carbon emissions and buy carbon offsets to neutralize their institutional emissions. For instance, The Sandbox computes carbon emissions from NFT transactions and acquires equivalent NRTs monthly, having purchased 5,854 NRTs at the time of writing.

Overall, buyers in the Nori market are grouped into four categories based on transactional behaviors, ranging from experiential purchases out of curiosity to voluntary carbon offsets for individuals and organizations. The diversity of buyers' motivation is consistent with the original design of the voluntary carbon offset market, which was to provide a broader trading space outside of the mandatory one. Moreover, individuals and small organizations, who are the primary buyers in the voluntary carbon offset market, do not have the capacity or incentive to build their own carbon offset systems, making a blockchain carbon offset market with third-party verification a trusted option for practicing carbon offsets. By leveraging the traceability feature of blockchain, we have been able to identify the composition of buyers in the voluntary carbon reduction market for the first time.

\subsection{Factors Affect Nori Market}

\begin{table*}[]    %格兰杰因果检验，没找到把ADF和gra结合的，如果表不够，或许在文字中描述各个var的ADF的t值
\centering
\renewcommand\arraystretch{1.3}   %表高度
% \caption{Granger causality test of potential influences on EUA and Nori markets}
\caption{Granger causality test of potential influences on EUA and Nori markets. The symbols '*', '**', and '***' indicate rejection of the null hypothesis at the 10\%, 5\%, and 1\% significance levels, respectively.}
\resizebox{\textwidth}{!}{    %自适应表宽度
\begin{tabular}{cccccccc}
\hline
\multirow{2}{*}{\textbf{Dependent variable}} & \multirow{2}{*}{\textbf{Independent variable}} & \multicolumn{2}{c}{\textbf{1 lag}} & \multicolumn{2}{c}{\textbf{2 lags}} & \multicolumn{2}{c}{\textbf{3 lags}} \\
                                    &                                       & \textbf{F value}     & \textbf{P value}     & \textbf{F value}      & \textbf{P value}     & \textbf{F value}     & \textbf{P value}      \\ \hline
\multirow{3}{*}{Nori}               & Energy Price                          & 0.3262      & 0.5680      & 1.1266       & 0.3247      & 5.1696      & 0.0015***    \\
                                    & Crypto Eco-prosperity                 & 0.3215      & 0.5709      & 0.2025       & 0.8167      & 0.2564      & 0.8568       \\
                                    & Public Opinions                       & 4.5562      & 0.0331**    & 2.2625       & 0.1048      & 1.6187      & 0.1836       \\ \hline
\multirow{3}{*}{EUA}                & Energy Price                          & 2.8563      & 0.0914*    & 1.6498       & 0.1928      & 1.0358      & 0.3760       \\
                                    & Crypto Eco-prosperity                 & 2.1383      & 0.1441      & 1.6739       & 0.1882      & 1.6682      & 0.1724       \\
                                    & Public Opinions                       & 0.9354      & 0.3338      & 0.5239       & 0.5924      & 0.3431      & 0.7942       \\ \hline
\end{tabular}
}
\label{tab:Gra}
\end{table*}

In this subsection, we aim to analyze the factors influencing the Nori market. We use the European Carbon Emission Allowance Futures (EUA), the largest mandatory market, as a comparison. We identify potential factors that may affect the market, collect relevant data, and test the causality between variables using the Granger causality approach in econometrics. We then present our analysis.
 
We begin by considering factors that may impact the carbon market and generate relevant proxy indices: Energy prices (using the price of IPE Rotterdam coal futures as a proxy), Crypto-ecological prosperity (using the token price-weighted index of the top 10 public chains in cryptocurrencies by market capitalization as a proxy), and Popular interest in blockchain-based carbon offset markets (using weighted results of Google search trend index for relevant keywords as a proxy). We then perform Granger causality tests between each pair of explanatory and response variables. Before performing the Granger causality test, we ensure each pair of variables passes the unit root test or confirms their cointegration relationship, to prevent creating pseudo-regressions. We test each variable up to the third-order lag, and the results are presented in Table~\ref{tab:Gra}.
 
Energy prices significantly impact both EUA (first-order lags) and Nori (third-order lags) markets. EUA's rapid response to energy prices is due to its efficiency from a long operational history and large trading scale. The primary traders on EUA are heavy industry companies sensitive to energy prices and can respond quickly to price changes. Comparatively, Nori, as a voluntary carbon emission reduction market, is also affected by energy prices. However, the market is less efficient due to its smaller size and emerging status. Additionally, according to the clustering analysis of buyers above, a significant portion of buyers (e.g., individual carbon offset users) are not price-sensitive. In summary, the Nori market's response to energy prices is slower.
 
The degree of crypto-ecological prosperity did not significantly impact the markets. Unlike most on-chain projects, Nori's trading has not been as strongly correlated with the crypto industry. In the three years Nori has been running, cryptocurrencies have gone through a bull and bear cycle, but Nori has not been affected too much. However, the opinion index has a significant impact on Nori, with a first-order lag. Reviewing Nori's development phase, we found a lot of marketing through public opinion. For example, Nori had launched Green NFT with The Sandbox, a virtual landscape of green plants, and the entire revenue generated from the purchase of this NFT was taken by The Sandbox to purchase NRTs to fuel carbon offsets. Similar marketings are Nori's main means of promoting itself and sustainability. Linking virtual assets with actual carbon offsets has proven effective in countering the negative perception of blockchain's energy consumption, with Nori emerging as a representative of environmentally friendly projects within the blockchain space.

In sum, the econometric tests identify two significant drivers for the Nori market: energy prices and public opinion. Specifically, the market is less sensitive to energy prices than the traditional mandatory carbon quota market but is more responsive to related public opinion. Hence, the trading of NRTs explores more potential for carbon credits.

\subsection{Social media opinion tracking}

Considering the significant impact of social interactions on the Nori market, we analyzed sentiment trends and topic frames of the Nori voluntary carbon offset market on Twitter. From December 2021 to December 2022, we randomly sampled one thousand tweets daily using the keyword "Nori." We used  Python library, TextBlob\footnote{https://github.com/sloria/TextBlob}, to calculate daily sentiment indices and used KH Coder\footnote{https://khcoder.net/en/} to obtain topic frames. For comparison, we also retrieved "NFT" related tweets.

\begin{figure*}[htbp]
    \centering
    \subfigure[]{
            \includegraphics[width=3.4in]{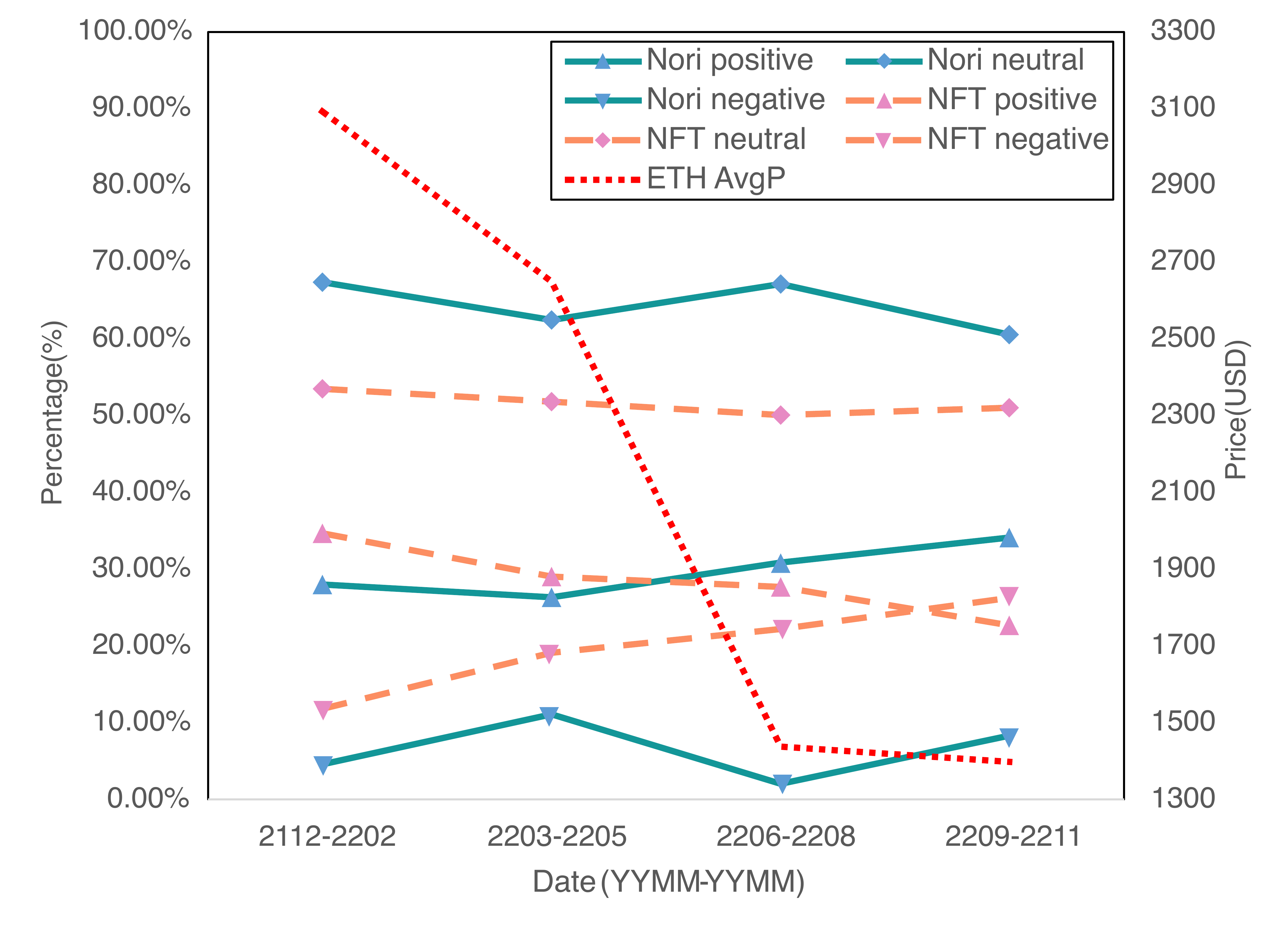}
    }
    \subfigure[]{
    	\includegraphics[width=3.4in]{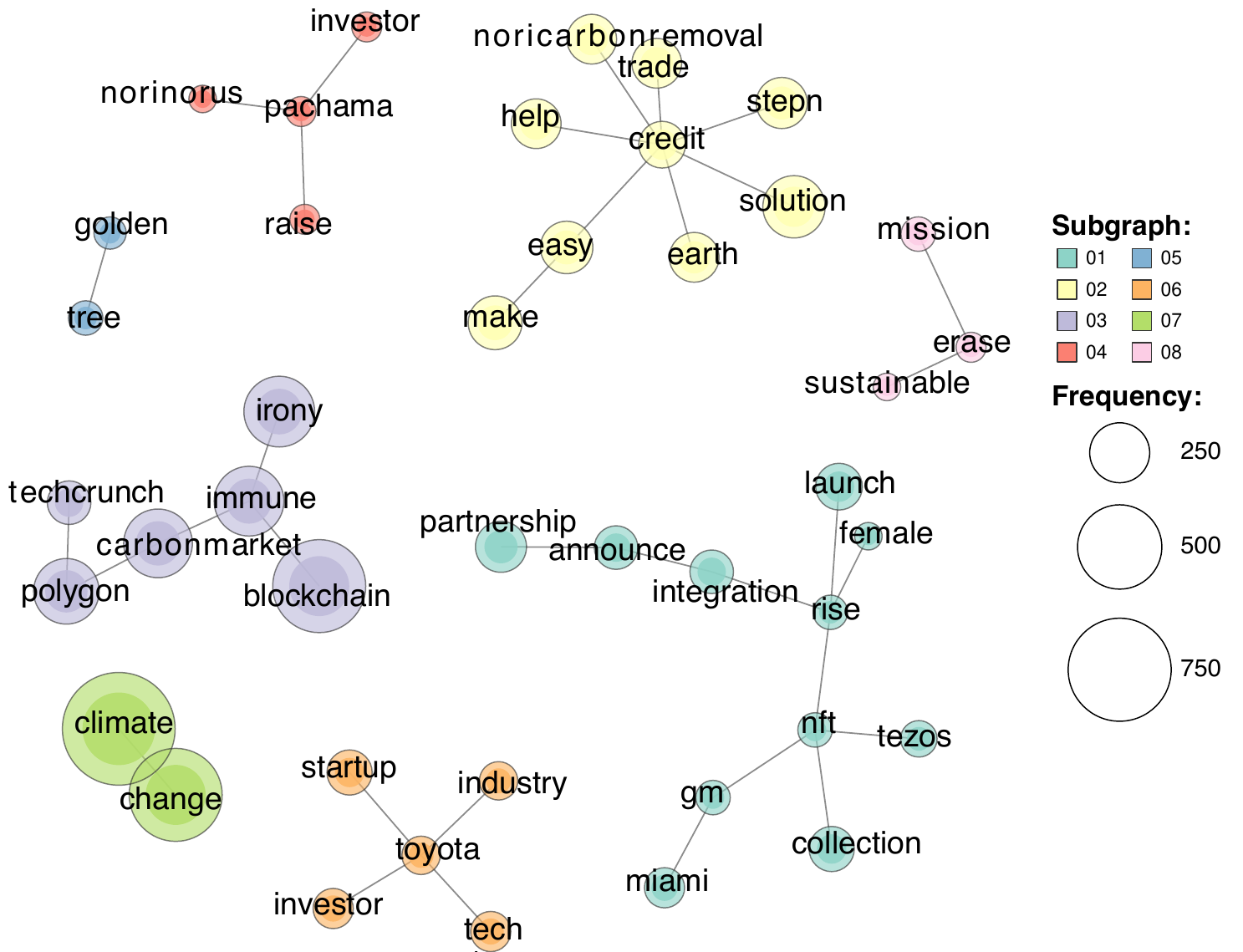}
    }
    \caption{Results of Nori carbon offset market's social media analysis. For subfigure (b), keywords were located based on the Fruchterman and Reingold algorithms, and correlations between keywords are calculated by the Jaccard coefficient and linked by edges. Each frame reveals a hot topic.}
    \label{social}
\end{figure*}

The sentiment analysis results are shown in Fig.~\ref{social}-A. It shows that since cryptocurrencies entered the bear market after 2022, the price of Ethereum (ETH), which is the cornerstone of decentralized applications, has been declining. In such an environment, the sentiment response to NFT was basically in line with the price trend of ETH, with positive sentiment gradually decreasing and negative sentiment gradually increasing, a result that belongs to the normal feedback of investors facing market fluctuations. However, the sentiment trend of Nori did not follow the overall fluctuation. For Nori, the positive and negative sentiment trend in 2022 is increasing, while the neutral sentiment is gradually decreasing. A communication event on Twitter often shows the following flow: a single campaign or influential communication node introduces Nori, followed by a subjective discussion of Nori by users who are not previously familiar with blockchain technology and industry dynamics, either positive or negative. The increase in positive and negative sentiment can be seen as the result of more subjective discussions by users who were not previously concerned about Nori and carbon offset, reflecting the utility of Nori's social media communication.

Fig.~\ref{social}-B displays the clustering of popular topics sampled from tweets over a year, revealing eight main topics with the highest buzz, each signifying an event or opinion related to Nori. Cyan subgraph (01) represents a collaboration in Miami between Nori and female NFT crypto artists on the Tezos blockchain, leveraging crypto social networks for carbon offset awareness. Yellow subgraph (02) covers discussions following STEPN's announcement to use Nori for offsetting its carbon footprint. Purple subgraph (03) depicts a blockchain project addressing carbon emissions by transitioning to the eco-friendly Polygon blockchain and purchasing carbon credits through Nori. Red subgraph (04) discusses the contrast between Pachama, an AI and satellite-driven carbon sequestration measurement project, and Nori. Blue subgraph (05) highlights the 'golden tree' NFT asset, jointly released by Nori and The Sandbox, with double the revenue from its sale allocated for CO2 removal. Orange subgraph (06) concerns Toyota Ventures' investment in Nori during a Series A funding round. Green (07) and pink (08) subgraphs discuss Nori's slogan "Reverse climate change with verified carbon removal," and the objectives of Nori’s carbon offset market. These topics reflect Nori’s diverse applications and partnerships, showcasing its expanding influence across various sectors.

In summary, we analyzed sentiment trends and hot topic frames to identify Nori's marketing and outcomes of related social interactions on social media. Nori's ongoing advocacy for sustainability and carbon offsets has profoundly impacted the crypto ecosystem. Over the past year, people's interest in Nori and carbon offset has grown. This rising trend is unaffected by broader crypto-market fluctuations and reflected by Nori's burgeoning partnerships and applications. Trending topics also show that some blockchain projects are already focusing on carbon emissions, switching their technical architecture to a less energy-intensive blockchain, and starting to implement voluntary carbon offsets.

\section{Conclusion}

This paper comprehensively examines blockchain's role in sustainability, with wireless technologies playing a pivotal role, through a resource lifecycle-based framework. It highlights blockchain's four utilities: recording and tracking, wide verification, value trading, and social interactions. Taking Nori as a case study, we explore its application in voluntary carbon offset markets, analyzing market participants, buyer types, and behavior patterns. Econometric causality testing helped identify market-influencing factors, while text mining shed light on trending topics and sentiments around Nori. Our findings indicate that Nori exemplifies how blockchain, bolstered by wireless technologies, facilitates the voluntary carbon market through the identified utilities. This paper suggests that Web3 native projects like Nori may significantly impact carbon offsetting and posits that sustainability and carbon offsets could emerge as a new consensus in the blockchain sector. Future research will focus on expanding the scope of blockchain technologies for carbon offsetting and identifying areas for contributing to sustainable development in Web3.

\bibliographystyle{IEEEtran}
\bibliography{IEEEabrv,rf.bib}

%\vspace{11pt}

\begin{IEEEbiographynophoto}{Chenyu Zhou} (chenyuzhou1@link.cuhk.edu.cn) is a M.Phil. student in computer and information engineering at The Chinese University of Hong Kong, Shenzhen, China. His research interests include blockchain, DeFi and Web3. Zhou received B.M. in Information Management and Information System, and B.Eng. in Computer Science and Technology from China University of Petroleum, Beijing, China.
\end{IEEEbiographynophoto}

\begin{IEEEbiographynophoto}{Hongzhou Chen} (hongzhouchen1@link.cuhk.edu.cn) is a Ph.D. student in computer and information engineering at The Chinese University of Hong Kong, Shenzhen, China, and a visiting student at Mohamed bin Zayed University of Artificial Intelligence, Abu Dhabi, United Arab Emirates. He also is a member of SeeDAO. His research interests include blockchain, DAO, DeFi and Web3. Chen received an M.A. in Global Studies from The Chinese University of Hong Kong, Shenzhen, China. He is a student member of IEEE.
\end{IEEEbiographynophoto}

\begin{IEEEbiographynophoto}{Shiman Wang} (shiman272@outlook.com) is a sustainable development consultant and a researcher in the Insight and Research Guild, SeeDAO. She received an M.A. in sustainability from The Chinese University of Hong Kong, Shenzhen, China. 
\end{IEEEbiographynophoto}

\begin{IEEEbiographynophoto}{Xinyao Sun} (asun@matrixlabs.org) is the co-founder and Chief Scientific Officer of Matrix Labs Inc., Vancouver, BC, Canada. He also holds the position of Postdoctoral Fellow and Tech Lead at the University of Alberta's Multimedia Research Center. In addition, he serves on the Web3 committee at the Canadian Blockchain Consortium. His Ph.D. research focused on using deep learning and signal processing techniques to improve the efficiency of satellite signal processing for wide-area monitoring in aerospace. Currently, his research focuses on web3-based digital worlds, smart contract automation, decentralized financial systems, and cross-chain security.

\end{IEEEbiographynophoto}

\begin{IEEEbiographynophoto}{Abdulmotaleb El Saddik} (elsaddik@uottawa.ca)  is the Acting Chair of the Computer Vision Department at Mohamed bin Zayed University of Artificial Intelligence, Abu Dhabi, United Arab Emirates, and Distinguished University Professor at the University of Ottawa, ON, Canada. His research interests include metaverse, digital twins, multimedia communication, connected health, and quality of experience. He is a Fellow of the Royal Society of Canada (FRSC), Fellow of the Engineering Institute of Canada (FEIC), Fellow of the Canadian Academy of Engineers (FCAE), and Fellow of IEEE (FIEEE).
\end{IEEEbiographynophoto}

\begin{IEEEbiographynophoto}{Wei Cai} (caiwei@cuhk.edu.cn) is an assistant professor of computer engineering at The Chinese University of Hong Kong, Shenzhen, China. He is also with the Shenzhen Institute of Artificial Intelligence and Robotics for Society. His research interests include decentralized systems and interactive multimedia. Cai received a Ph.D. in electrical and computer engineering from The University of British Columbia (UBC), Vancouver, Canada. He is a senior member of IEEE.
\end{IEEEbiographynophoto}

\end{document}